\documentclass[runningheads]{llncs}
\usepackage[T1]{fontenc}
\usepackage{graphicx}
\usepackage{url}
\usepackage{color}
\usepackage{caption}
\usepackage{subcaption}

\begin{document}
\title{Analyzing I/O Performance of a Hierarchical HPC Storage System for Distributed Deep Learning}
%
%
\author{Takaaki Fukai\inst{1}\orcidID{0000-0003-4216-4807} \and
Kento Sato\inst{2}\and
Takahiro Hirofuchi\inst{1}\orcidID{0000-0002-1253-6625}}
\authorrunning{T. Fukai et al.}
%
\institute{National Institute of Advanced Industrial Science and Technology (AIST), Tokyo, Japan\\
\email{\{takaaki.fukai, t.hirofuchi\}@aist.go.jp} \and
RIKEN Center for Computational Science, Kobe, Japan\\
\email{kento.sato@riken.jp}}
\maketitle              
\begin{abstract}
Today, deep learning is an essential technology for our life.
To solve more complex problems with deep learning, both sizes of training datasets and neural networks are increasing.
To train a model with large datasets and networks, distributed deep neural network (DDNN) training technique is necessary.
For large-scale DDNN training, HPC clusters are a promising computation environment.
In large-scale DDNN on HPC clusters, I/O performance is critical because it is becoming a bottleneck.
Most flagship-class HPC clusters have hierarchical storage systems.
For designing future HPC storage systems, it is necessary to quantify the performance improvement effect of the hierarchical storage system on the workloads.
This paper demonstrates the quantitative performance analysis of the hierarchical storage system for DDNN workload in a flagship-class supercomputer.
Our analysis shows how much performance improvement and volume increment of the storage will be required to meet the performance goal.


\end{abstract}

\section{Introduction}
\label{sec:Introduction}
Today, the demand for large-scale deep learning has significantly increased.
The sizes of training models and datasets for the training are expanding to meet the demand.
For example, training models for image classification, such as EfficientNet~\cite{Tan2019_EfficientNet} with large datasets, such as OpenImages dataset~\cite{Kuznetsova2020_OpenImages} are used.
Some computational science applications also use deep learning methods such as CosmoFlow~\cite{Mathuriya2018_CosmoFlow} and DeepCam~\cite{kurth2018_DeepCam}.
A single machine does not have enough computational and memory capacity for these large workloads. 
Therefore, distributed deep neural network (DDNN) training technique, which allows training models on multiple machines connected by a network, is necessary.
HPC, optimized for huge and distributed workloads, are promising environments for large training workloads.

I/O is becoming a bottleneck in the training workloads for the following reasons~\cite{Mohan2021,Pumma2019}.
The first reason is dataset growth.
The size of datasets is increasing for training higher quality models~\cite{Mahajan2018,Beal2022}.
Large datasets that do not fit in memory cause training applications to issue many I/O requests.
The second reason is expanding performance gap between computation and I/O.
Although computation time is becoming shorter by distributed execution techniques, I/O performance is not improved.
This expands the performance gap.
Therefore the I/O performance of future HPC clusters is crucial.

For designing a storage system for future HPC, it is crucial but difficult to find out what improvement of storage systems would achieve a performance goal of a target I/O intensive DDNN workload.
The first reason why it is difficult is that most storage systems in flagship-class HPC clusters are hierarchical, which combines fast but small storage, and a slow but large storage system~\cite{Sato2020_Fugaku,Vazhkudai2018_Summit}.
Therefore, it is unclear which we should pay our cost for the throughput of the global or local filesystems or the volume for the local file system.
The second reason is that tuning a DNN application for distributed execution requires several weeks or months for a new cluster or processor architecture.
Therefore, we need a much cost and time to find out the I/O bottleneck in a DDNN training workload.

This paper shows a case study on the performance analysis of a hierarchical storage system for DDNN workload and the estimation of necessary improvement of the storage systems to meet a performance goal.
To reveal the effect of faster storage, we first measure the I/O operations time of a synthetic I/O intensive training workload with various proportions of fast and slow storage sizes.
Then we estimate the impact of storage system improvement on the training performance based on a result of the I/O performance analysis.
The method can estimate the contribution of various improvement of a hierarchical storage system, to overall the training performance.

The contributions of this work are: (1) A methodology to study the I/O bottleneck of DDNN training workloads in the hierarchical storage system; (2) A methodology to explore options for improvement of a storage system to achieve a performance goal of DDNN training workloads.

The remainder of this paper is organized as follows.
Section~\ref{sec:Background} explains the background.
Section~\ref{sec:RelatedWork} reviews the related work.
Section~\ref{sec:Methodology} explains our method.
Section~\ref{sec:Evaluation} demonstrates our methods on a flagship-class supercomputer.
Section~\ref{sec:Discussion} discusses the potential of the method.
Conclusions are presented in Section~\ref{sec:Conclusion}.

\section{Background}
\label{sec:Background}

\subsection{File access in distributed neural network workloads}
The file access pattern by DNN training applications is different from that in scientific computational applications.
In training, stochastic gradient descent (SGD) is a common technique to improve training speed and accuracy\cite{Akiba2017_ResNet50_15min,Mikami2018_ResNet50_122sec,Yamazaki2019_ResNet50_74.7sec}.
In SGD, a program splits a training dataset into mini-batch and inputs a mini-batch to the neural network.
To avoid the degradation of training accuracy due to a fixed input order, it shuffles the order of the files of the dataset every time when inputting all samples to the neural network.
Therefore the program accesses each file once an epoch in a random order.
It is hard to apply general cache policies because of less time and spatial locality.
If the dataset is larger than the memory volume for page caches, the cache miss on reading file often occurs.
It is a reason why I/O is easy to be the bottleneck.

In distributed training, multiple compute nodes read the dataset simultaneously.
In data-parallel, which is one of common parallelizing techniques, each compute node has a part of the dataset and calculates it.
There are two data shuffling manners called local shuffling and global shuffling.
Local shuffle means that each process only shuffles and reads a part of the dataset.
On the other hand, global shuffle means that the application shuffles the whole dataset and splits it for each computer every epoch.
Local shuffling is easy to use local storage for each computer because the computer needs to access only the initial allocated part of the dataset.
However, it reduces the training accuracy because it reduces the randomness of the input dataset.
In some cases, the local shuffle approach is not suitable due to the accuracy degradation.
On the other hand, the global shuffling does not affect the training accuracy, but replacing the part of the dataset for each epoch is a heavy I/O workload, especially, training with a large dataset and a large number of computers.
Therefore, we focus on the global shuffling in this paper.

\subsection{Storage system in HPC}

Recent flagship class HPC clusters provide a hierarchical storage system.
Hierarchical storage systems typically consist of a small but fast storage system and a large but slow storage system.
HPC clusters often provide the former as a local file system (LFS) and the latter as a global file system (GFS).
Summit \cite{Vazhkudai2018_Summit} provides node-local burst buffers (node-local NVMe SSD) and a parallel file system (IBM’s SpectrumScale GPFS\texttrademark).
Fugaku\cite{Sato2020_Fugaku} also provides a hierarchical storage system that consists of the 1st level storage (an SSD for every 16 nodes) and the 2nd level storage (a Global storage system).
We assume that DDNN applications in a global shuffle manner use the local storage in the hierarchical storage as the cache of global storage.
An important question to answer to design future storage systems in HPC for machine learning workload is the best balance of fast and slow storage from the viewpoint of size and performance.

\section{Related work}
\label{sec:RelatedWork}

There are several works to analyze and model the DNN performance.
Wang et al. proposed a modeling method for the DNN training workload based on the Roofline model~\cite{Wang2020_RooflineForDL}.
They focus on the performance of the computation and memory accesses however the I/O performance is not considered. 

There are several works for analyzing and optimizing I/O performance for DDNN workloads.
Several works~\cite{Pumma2019,Mohan2021} have analyzed the I/O performance for the DDNN and proposed optimization methods.
Devarajan et al. proposed a benchmark to measure the I/O performance for DDNN and find the opportunity for tuning I/O parameters~\cite{Devarajan2021_DLIO,DevarajanGithub_DLIO}.
These works assume a non-hierarchical storage system.
We focus on I/O performance of hierarchical storage systems.

Several works~\cite{Zhu2018_DeepIO,Serizawa2019,Zhu2019_DLFS,Dryden2021_ClairvoyantPrefetching} assume hierarchical storage systems in their I/O optimization method for DDNN workloads.
They focus on application-level optimization to solve the I/O bottleneck.
On the other hand, our work is toward performance improvement of storage systems.

Paul et al. analyzed the I/O log generated by all the jobs on Supercomputer Summit during a year~\cite{Paul2021}. 
They revealed the tendency of ML jobs and the usage of the storage system by them, especially, the usage of the burst buffer.
In this work, the 23,389 ML jobs of 845,036 jobs in 2020 on Summit were analyzed.
The analysis results suggested a rapid increment in the use of ML technologies in HPC, and some of the ML jobs used the burst buffer in addition to the GPFS.
This work analyzes the comprehensive analysis of the real ML workload from the viewpoint of usage of the hierarchical storage system in the HPC environment.
Our work analyzes the I/O performance of hierarchical storage systems in detail.

\section{Methodology}
\label{sec:Methodology}

\subsection{Overview}
\label{sec:Methodology_Overview}
Our analysis method is composed of three steps, (1) measuring I/O performance, (2) analyzing measurement results, and (3) estimating the impact of the speed-up of global and local storage on training performance.

In the measurement, we execute a DDNN benchmark with profiling the I/O on a hierarchical storage system.
The benchmark reads a dataset from the hierarchical storage system and uses LFS as the cache of GFS.
To reveal how LFS contributes to overall the I/O performance, we measure the I/O performance with various proportions of the size of the cached data on LFS.
We expect that the performance characteristics depend on a performance balance of GFS and LFS as well as the sizes of files in a dataset.
Therefore, we measure the I/O performance with multiple performance balances and the file sizes combination.

In the analyzing step, we analyze the profiling data separately by file system and by type of I/O operations.
To do this, we break down the I/O time into the following four I/O classes based on the I/O profiling data obtained in the benchmark execution.
\textit{GFS-READ} is a class for read operations on a GFS,
\textit{GFS-META} is a class for metadata operations (\verb|open()|, \verb|close()|) on a GFS,
\textit{LFS-READ} is a class for read operations on an LFS,
and \textit{LFS-META} is a class for metadata operations on an LFS.
Note that file operations on the dataset in a DNN training are only \verb|open()|, \verb|close()|, and \verb|read()| because applications do not make any modifications and new samples.
We target the I/O time of the slowest process among all parallel processes, because it is the most dominant for the overall training time.

In the estimating step, we extrapolate from the above results, expected training time enabled by the speed-up of global and local storage.
We first calculate the expected overall I/O operation time on the assumption that the speed of an I/O class is improved by a given ratio.
We also calculate the expected impact on training time by the performance improvement of multiple I/O classes and which combination of the improvement will satisfy the performance goal.

\subsection{Measuring I/O performance by benchmark}
\label{sec:Methodology_Measurement}
To measure the I/O performance, we use DLIO~\cite{Devarajan2021_DLIO} benchmark, a benchmark for I/O performance on distributed deep neural network workloads.
DLIO benchmark supports distributed execution and generating the synthetic dataset for the benchmark.
However, it does not support hierarchical storage.
Therefore, we add the three functions to DLIO for our measurement of hierarchical storage systems.
The functions are (1) Reading the dataset from both GFS and LFS with a specified proportion,
(2) Global shuffling, and 
(3) Generating the synthetic files on the local filesystem by each compute node.

In the benchmark execution, the cached files are not evicted, in other words, the cache policy is pinning.
As described in Section~\ref{sec:Background}, the training application accesses all samples the same number of times.
Therefore, the cache hit rate with the pinning policy is the same as the percentage of the cached file~\cite{Mohan2021}.

We prepare two datasets, a small file dataset and a large file emulating ImageNet dataset and CosmoFlow dataset respectively.
The small file dataset consists of 128 KiB files and the large file dataset consists of 12 MiB files.
The numbers of files in the small and large datasets are 589824 and 6144 respectively.
The total size of both datasets is 72 GiB so that whole of the dataset can be on the LFS and all of the processes read the same number of files.
Because the entire dataset cannot be put on the memory of each compute node (32 GiB), the benchmark application reads the files from the filesystem.
The file format in both datasets is tfrecord, and the number of samples in each file is one.

To measure the I/O performance with multiple performance balances of the GFS and LFS, we measure the I/O performance with different numbers of the object storage targets (OSTs) of the lustre-based GFS.
We can limit the number of OSTs to 1 by \verb|lfs| command.
Therefore, we measure the I/O performance with all provided OSTs (faster GFS) and 1 OST (slower GFS).

To measure the I/O performance in the benchmark execution, we use Darshan~\cite{Darshan}, which is a profiling tool for I/O.
Darshan can capture and record each file operations such as \verb|open()|, \verb|close()|, and \verb|read()|.

\subsection{Analyzing the I/O performance}
To reveal the bottleneck in detail, we break down the I/O time of the slowest process into the following four I/O classes and recognize which I/O class is a bottleneck.
We calculate the I/O time for each process and find the slowest one, which dominates the training performance.
We analyze the I/O performance from the log generated by darshan using \verb|darshan-parser| command \cite{Darshan-util}.
We calculate for each I/O time based on \verb|POSIX_F_READ_TIME| and \verb|POSIX_F_META_TIME|.

\subsection{Estimate performance by storage improvement}
\label{sec:Methodology_Estimation}
To estimate impact of $N$\% throughput improvement of a I/O class, we calculate $\times \frac{100}{100+N}$ of the measured time of the I/O class.
The improvement may not directly affect the total I/O time because the improvement may change the bottleneck to another I/O class.
Therefore, we calculate total I/O time for each process with the improvement of a class, then pick the slowest process.

\section{Experiment results}
\label{sec:Evaluation}

\subsection{Setup for experiment}
We perform the experiments on Supercomputer Fugaku\cite{Sato2020_Fugaku}.
Compute nodes of Fugaku has 48 computing cores of A64FX and 32 GiB HBM2 memory.
Fugaku has a hierarchical filesystem comprising the 1st- and 2nd-level filesystems named LLIO and FEFS, respectively.
In our measurement, we regard the LLIO as a local filesystem (LFS) and the FEFS as a global filesystem (GFS).
FEFS is a lustre-based parallel filesystem and it has 60 OSTs in Fugaku.
One per 192 compute nodes connected to the FEFS by InfiniBand EDR.
The other compute nodes connected by TofuD access to FEFS via the network and the compute node.
For LLIO, one per 16 compute nodes has an NVMe SSD and the other compute nodes connected access to the SSD via the network and the compute node.
LLIO provides three areas, node temporary area, shared temporary area, and 2nd-layer cache area\cite{Akimoto2020_FugakuTechReview}.
In our measurement, we only use node temporary areas, which is a dedicated area for a compute node, because the transparent cache does not allow us to control caching files of the dataset on LLIO.

In our measurement, we run the DLIO benchmark on the 768 compute nodes of Supercomputer Fugaku as batch jobs.
The four processes execute on every compute nodes, so that total number of processes is 3072.
The node layout is $8 \times 6 \times 16$ in the TofuD torus network.
We also pass an option to the job scheduler to strict the position of the node connected to the GFS.

Before executing the benchmark job, we generate the datasets on the GFS.
Because the system removes data on LFS after finishing the job, every benchmark job generates the same dataset on LFS as GFS.
Note that the job generates the dataset instead of copying the dataset from GFS to reduce the setup time.

We execute the benchmark with every 5\% from 0\% to 100\% cache rate.
We set the calculation time in the DLIO to zero.
Therefore, the DLIO reports only the I/O and data processing time.
The number of epochs is three to avoid making the darshan log files huge.
The prefetch of the dataloader is enabled so that it is not synchronized for each iteration even if the computation threads are synchronized for all-reduce communication.
The batch size is 12 for the small file dataset and 2 for the large file dataset.

\subsection{Measuring execution time for epochs}

In our experiments, we first measure the execution time of the DLIO benchmark and I/O time in the benchmark execution with various settings as mentioned in \ref{sec:Methodology_Measurement}.
We reveal the difference in the performance depending on file sizes in the datasets and speeds of GFS.
Additionally, by comparing the execution time reported by the benchmark and the total I/O time reported by the I/O profiler, we verify that the benchmark is I/O intensive.

\begin{figure}[t]
  \includegraphics[width=\textwidth]{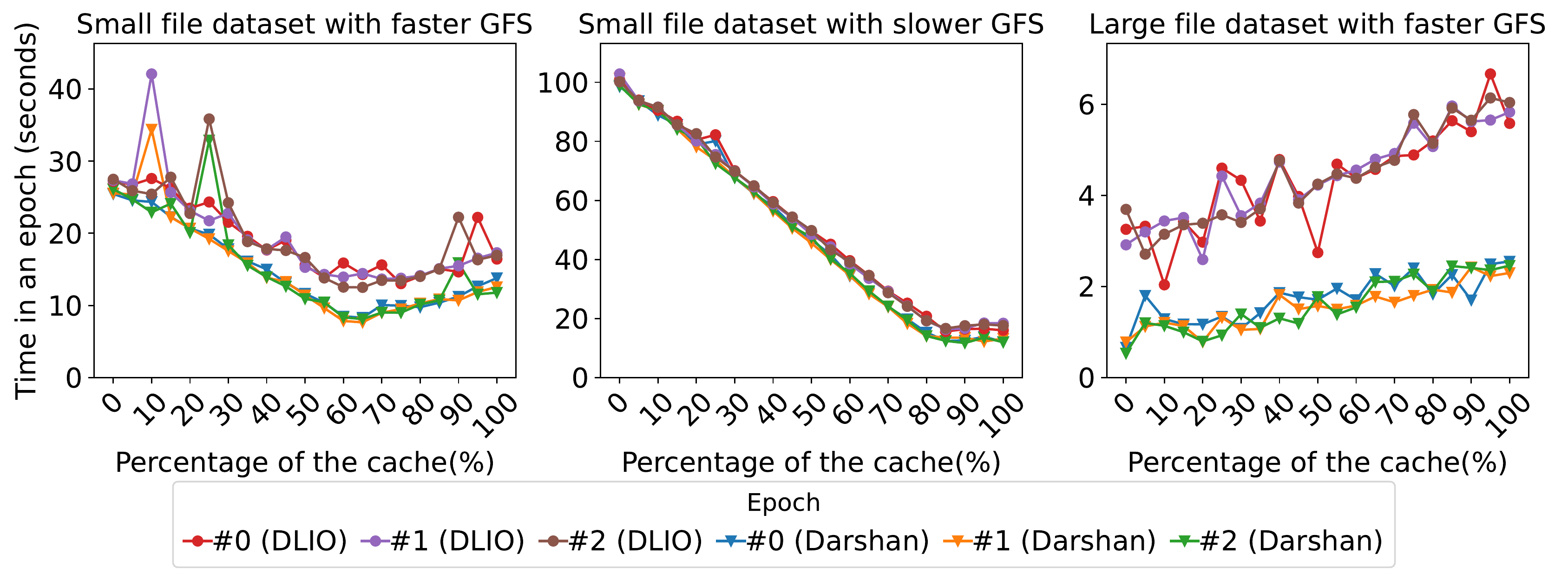}
  \caption{Execution time of DLIO benchmark and I/O time reported by Darshan}
  \label{fig:epochtime_vs_iotime}
\end{figure}

Figure \ref{fig:epochtime_vs_iotime} shows the execution time and I/O time for each epoch in a job.
The x-axes of the graphs show the percentage of the files on the LFS.
The y-axes show the execution time of an epoch.
The graph shows the results of 3 epochs in a benchmark execution.
The lines with round markers are the execution time reported by the DLIO benchmark, and those with triangular markers are the I/O time reported by Darshan.
Because the I/O time of the slowest I/O process is dominant, we calculate and plot them as I/O time on the graph.

The results show that the impact of the LFS on the training performance depends on the file sizes and the performance balance of GFS and LFS.
The effect of the LFS with the 1 OST of GFS is larger than that with the 60 OSTs.
The reason is the performance difference between LFS and GFS on the 1 OST is larger than that on 60 OSTs.
In 12 MiB files workload with 60 OST of GFS, the LFS does not contribute to the performance improvement, and using only the GFS with the 60 OST is the best.

About the I/O time, the graph indicates that the execution time is constantly longer about 2 sec, but it is strongly related to the I/O time.
This result indicate that the I/O time of the slowest process during the synchronizations among the processes strongly interrelates to the training performance.

\subsection{Analyzing I/O performance}

\begin{figure}[t]
  \includegraphics[width=\textwidth]{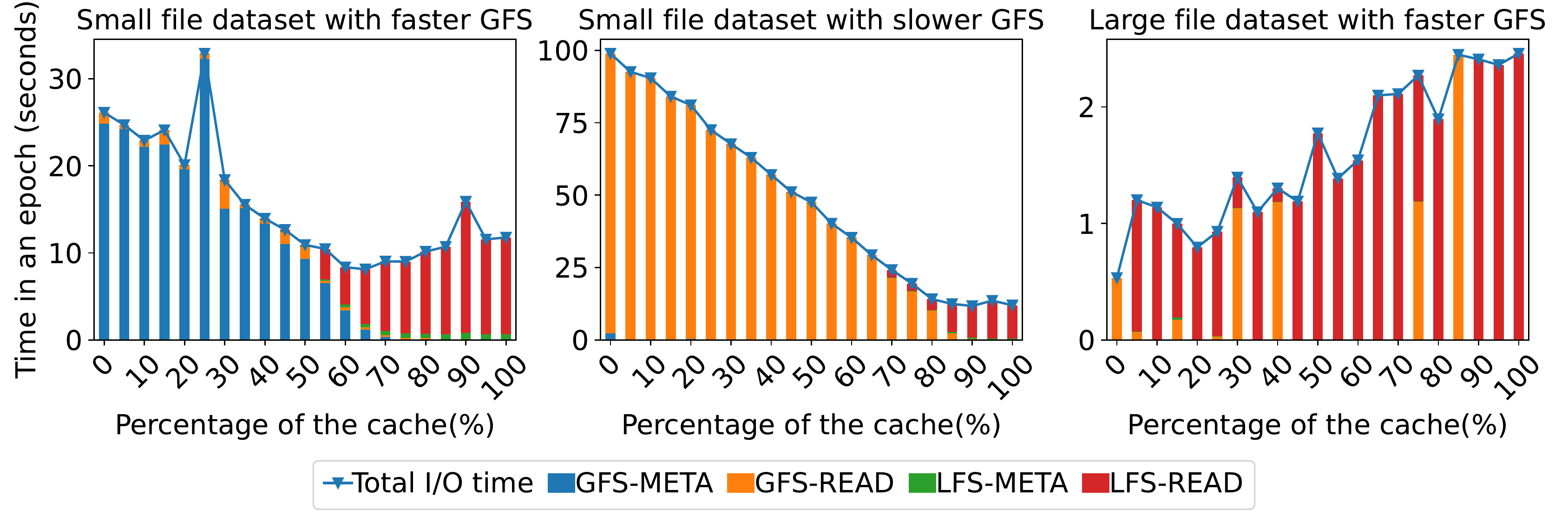}
  \caption{Break down the I/O time of the slowest process for each epoch (Note: Range of y-axis are different)}
  \label{fig:iotime-per-op}
\end{figure}

Next, we classify the Darshan records into the four I/O classes and calculate the I/O time for each class.
Figure~\ref{fig:iotime-per-op} shows the result of the breakdown of the \#2 epoch in the previous graphs.
Each line shows the total I/O time same as Figure~\ref{fig:epochtime_vs_iotime}.
According to the result, the bottleneck is different depending on the setup.

In 128 KiB files workload with the faster GFS, the bottleneck is GFS-META when less than 60\% of data is on the LFS.
On the other hand, the bottleneck is changed to LFS-READ with more than 60\% cached data on LFS.
We think when more than 60\% of data is put on LFS, LFS throughput is saturated.
Therefore, the read time from LFS is linearly increased with the percentage of the cached data.
So putting more than 60\% of data on the cache in the workload does not contribute to the training performance.
For example, in training with ImageNet dataset whose size is almost 150 GB, almost the 90 GiB LFS for each compute node is enough to achieve the best I/O performance by the hierarchical storage system.
With the 60\% cached data, both GFS-META and LFS-READ are included in the I/O time of the slowest process.

As compared with the faster GFS, the I/O bottleneck in the workload with the slower GFS is much different.
The graph in the middle of Figure \ref{fig:iotime-per-op} shows that the bottleneck is GFS-READ instead of GFS-META with small percentages of the LFS (less than 80\%).
We think that the reason why GFS-META time becomes shorter is reducing the load on the metadata server of FEFS due to the lower throughput of the GFS.

As compared with the small files workload, the I/O bottleneck in the large files workload with the faster GFS is also much different.
The right side graph in Figure \ref{fig:iotime-per-op} shows that the bottleneck is the LFS-READ in most of the cases.
Because the number of metadata operations is much smaller than that in the small file workload, the GFS fully provides its bandwidth without the bottleneck by the metadata operation.
As a result, the total bandwidth of the GFS is higher than LFS.
It means that the number of compute nodes is not enough to take advantage of the scalability of the LFS.
Note that the 768 nodes are not so large scale as a workload in Fugaku, however from viewpoint of the machine learning workload, the number of nodes is large enough to lead to a large batch problem.

From viewpoint of exploration of storage design for a performance goal, the result on small file dataset and faster GFS (the left side graph in Figure \ref{fig:iotime-per-op}) is challenging situation because multiple I/O class is included in the I/O time in the fastest result (cache rate = 65\%).
This means that improving only one I/O class processing will not be enough to improve entire I/O performance.
Therefore, we pick up the result to demonstrate our estimation method of the I/O improvement effect.

\subsection{Estimating the impact of the storage improvement}

\begin{figure}[t]
  \begin{subfigure}{0.49\hsize}
    \includegraphics[width=\textwidth]{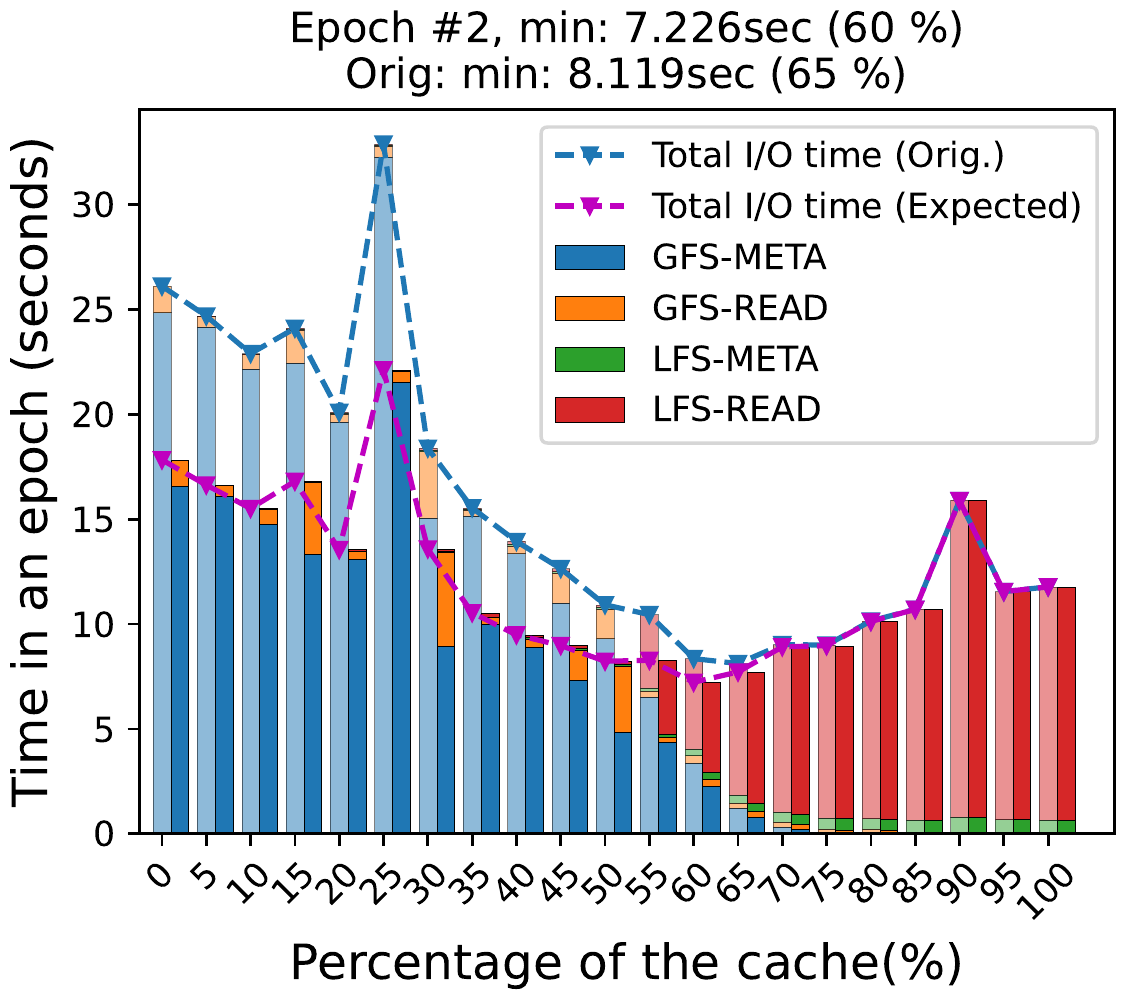}
    \caption{Improving GFS-META}
    \label{fig:expected-iotime-per-op-128KiB-60OST-50P-faster-left}
  \end{subfigure}
  \begin{subfigure}{0.49\hsize}
    \includegraphics[width=\textwidth]{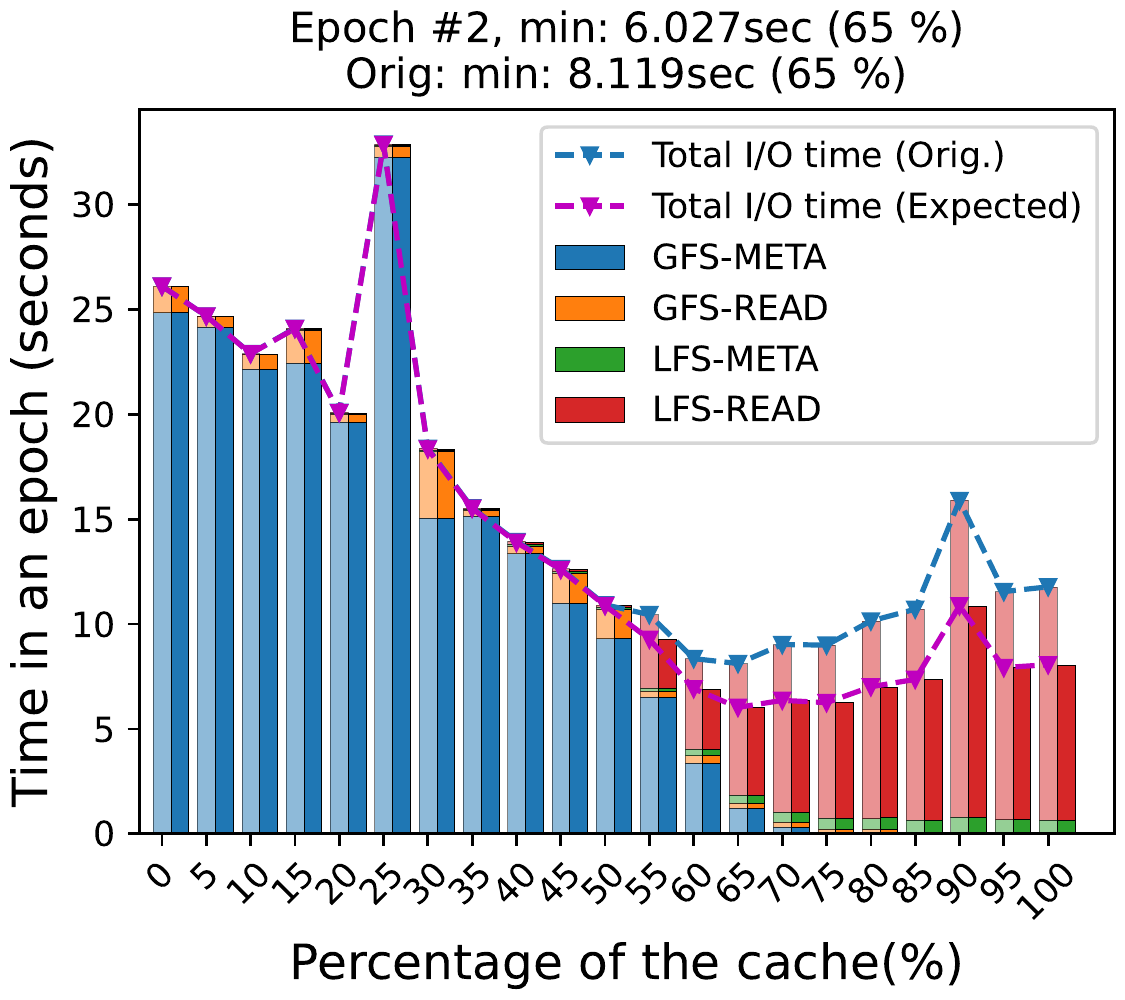}
    \caption{Improving LFS-READ}
    \label{fig:expected-iotime-per-op-128KiB-60OST-50P-faster-right}
  \end{subfigure}
  \caption{Expectation the I/O time with 50\% improvement with small file dataset and 60 OSTs of the GFS}
  \label{fig:expected-iotime-per-op-128KiB-60OST-50P-faster}
\end{figure}

As mentioned in \ref{sec:Methodology_Estimation}, we estimate the performance improvement by simple calculation based on the analysis result.
Figure~\ref{fig:expected-iotime-per-op-128KiB-60OST-50P-faster} shows the result of the estimation of the impact of a 50\% improvement of GFS-META (Figure~\ref{fig:expected-iotime-per-op-128KiB-60OST-50P-faster-left}) and LFS-READ (Figure~\ref{fig:expected-iotime-per-op-128KiB-60OST-50P-faster-right}).
The axes in the graph are the same as in Figure~\ref{fig:iotime-per-op}.

Figure \ref{fig:expected-iotime-per-op-128KiB-60OST-50P-faster-left} shows the estimation result of improving the GFS-META by 50\%.
The best combination of the GFS and LFS is changed from 65\% to 60\% LFS, and the slowest I/O time is reduced by almost 12.8\% in the best case.
Figure \ref{fig:expected-iotime-per-op-128KiB-60OST-50P-faster-left} shows the estimation result of improving the LFS-READ by 50\%.
The best cache rate is not changed, and the slowest I/O time in the best case is reduced by 24\%.

\begin{figure}[t]
  \begin{center}
    \includegraphics[width=0.5\textwidth]{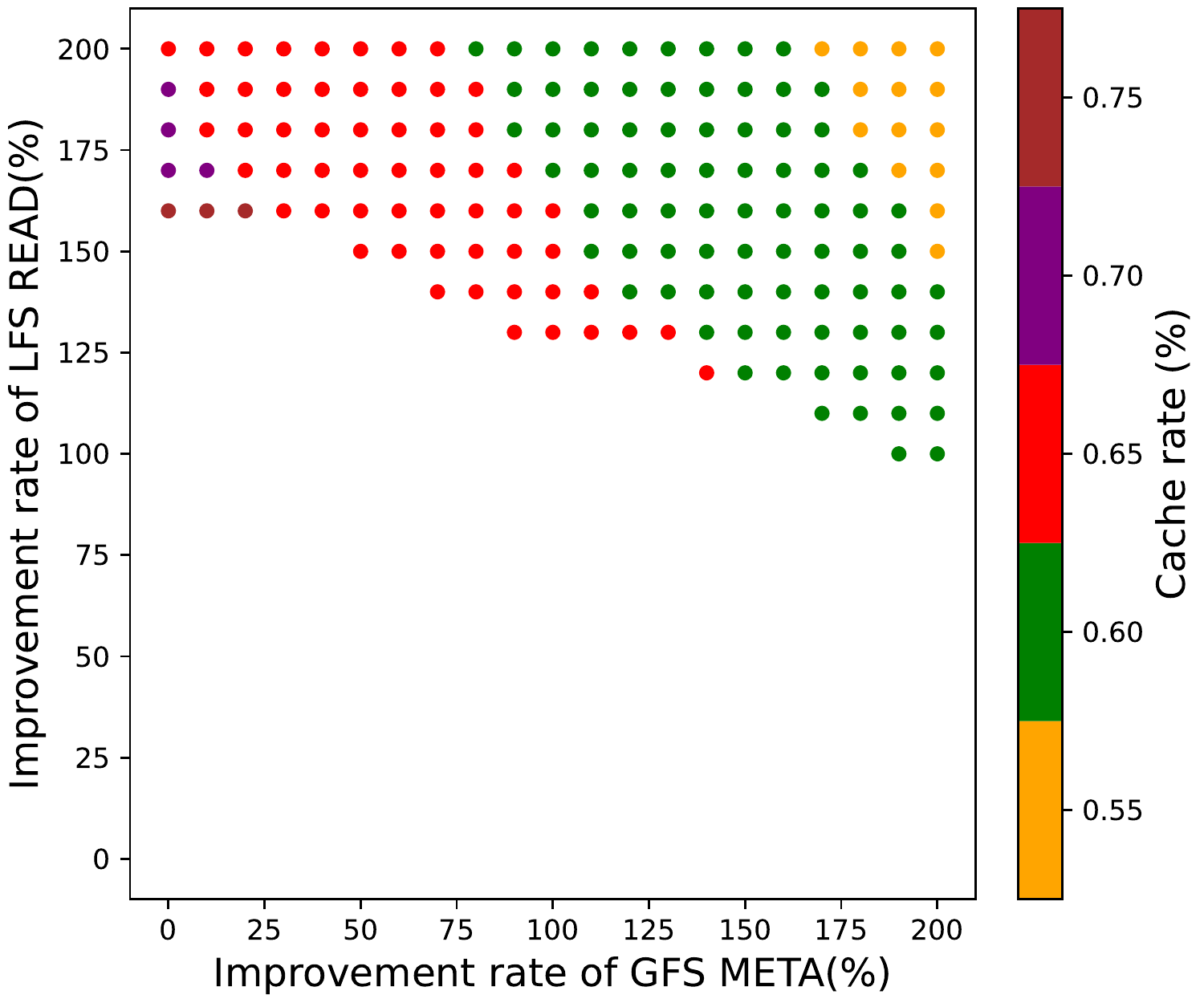}
  \end{center}
  \caption{The estimation of performance improvement of GFS-META and LFS-META for meeting performance goal of 4 sec / epoch (128 KiB, 60 OST GFS)}
  \label{fig:two-improvement-for-iotime-goal-4sec-128KiB-60OST}
\end{figure}

Next, we show the estimation of the impact of improvement of two operations classes simultaneously.
There are many parameters and values such as improvement rate for each operation, cache rate, and the I/O time.
All of them are too many to put on a single graph.
Again, the architect of the system needs knowledge of the given performance goal.
Therefore, we show the estimation by indicating which improvement combination would meet the performance goal.

Figure \ref{fig:two-improvement-for-iotime-goal-4sec-128KiB-60OST} shows the sufficient combinations of the performance improvement on two classes, GFS-META and LFS-READ, on the small files dataset and the faster GFS workload.
The result in the graph is based on the measurement of I/O time in the \#2 epoch.
The x and y axes show each improvement rate.
On the graph, the dot is plotted if the improvement combination will meet the given performance goal.
The graph indicates a result for the performance goal of 4 seconds I/O time in an epoch.
Additionally, the colors of the dots indicate the minimum cache rate to meet the goal.
For example, to achieve 4 seconds I/O time in an epoch with a 65\% cache rate, at least 120\% improvement of LFS-READ is required.
In that case, a 140\% improvement of the GFS-META is required.
The architect can explore the option of the improvement choice by the plot.

\section{Discussion}
\label{sec:Discussion}

In our evaluation, we assume the global shuffling manner to exploit GFS.
However, training applications with the local shuffle also can combine the LFS and GFS to put larger chunks of the dataset than that with only the LFS.
In this case, the size of the LFS and the randomness of the shuffling are a trade-off.
To consider how large chunks are preferred, the application user also can use our method to find the contributions to the performance of the LFS.

In our evaluation, we assume a pinning cache policy on the LFS.
However, our analysis can apply to the other cache policy if the cache hit rate can be calculated.
You can replace the "cache rate" with "cache hit rate" in the analysis result because both are the same in DNN workloads with the pinning policy. Then you can find the required size of the LFS from the relation between the cache hit rate and the size of the cache in your better cache policy.

In our evaluation, we estimate the improvement by a simple calculation.
However, the performance characteristic may not be simple.
For example, the estimation from the measurement results with 1 OST of the GFS with the simple calculation does not fit that with the 60 OSTs of GFS.
For more accurate estimation, improving the calculation method is necessary by modeling the characteristic.
The considerable approach is based on machine learning or queueing theory.
Even if the calculation method will be improved, our plot method shown in Figure \ref{fig:two-improvement-for-iotime-goal-4sec-128KiB-60OST} is useful for the storage system architect.

\section{Conclusion}
\label{sec:Conclusion}
This paper presented a case study on the performance analysis of a hierarchical storage system for a DDNN workload in a flagship-class HPC cluster, discussing potential performance improvement enabled by the speed-up of the storage system.
We also estimated the improvement of training performance by various improvement of the hierarchical filesystem.
The analysis result showed that the I/O bottleneck in the training workload depends on performance balance between global and local storage as well as file sizes in a dataset.

Our estimation showed that the performance improvement of a global filesystem will contribute to reducing the necessary volume size of a local filesystem, and the performance improvement of the local file system will contribute to reducing fastest I/O time.
Our estimation method can help architects of HPC filesystems to find the necessary performance and the volume size of the local and global filesystems to meet a given performance goal.

Because our proposed method needs the measurement of I/O performance at least once, one of our future works is exploring a simpler or no measurement-required method.
The other future work is to build the performance modeling of the storage system for more accurate estimation.

\bibliographystyle{splncs04}
\bibliography{paper-pdcat2022}

\end{document}